\UseRawInputEncoding
\documentclass[twocolumn,floatfix,nofootinbib,showpacs]{revtex4-1}
\usepackage[margin=0.75in]{geometry}

\usepackage{graphicx,amsfonts,amssymb,amsmath,mathrsfs,bm,dsfont}
\usepackage[normalem]{ulem}
\usepackage[title]{appendix}
\usepackage{commath}
\usepackage{mathtools}
\usepackage{xcolor}         
\usepackage{array} 
\newcolumntype{H}{>{\iffalse}c<{\fi}@{}}
\usepackage{lipsum}
\usepackage{booktabs}
\usepackage{bm}
\usepackage{amsthm}
\usepackage{algpseudocode}
\usepackage[utf8]{inputenc}
\usepackage{physics}
\usepackage{amsfonts}
\usepackage[ruled,vlined,algo2e]{algorithm2e}
\usepackage{algorithm}
\usepackage{bbm}
\usepackage{color}
\usepackage{multirow}
\usepackage{hhline}
\usepackage{parskip}

\def\ket#1{\vert#1\rangle}

\def\Longarrow{\protect\@lra}
\def\@lra{\relbar\joinrel\relbar\joinrel\relbar\joinrel%
          \relbar\joinrel\rightarrow}

\usepackage[colorlinks=true, breaklinks=true, linkcolor=blue, citecolor=blue, urlcolor=blue]{hyperref}

\begin{document}

\title{Financial Risk Management on a Neutral Atom Quantum Processor}

\author{Lucas Leclerc$^{1,2}$}
\thanks{These authors contributed equally to this work}
\author{Luis Ortiz-Gutiérrez$^1$}
\author{Sebastián Grijalva$^1$}
\author{Boris Albrecht$^1$}
\author{Julia R. K. Cline$^1$}
\author{Vincent E. Elfving$^1$}
\author{Adrien Signoles$^1$}
\author{Loïc Henriet$^1$}
\email{loic@pasqal.com}
\affiliation{$^1$PASQAL, 7 rue Léonard de Vinci, 91300 Massy, France}
\affiliation{$^2$Universit\'e Paris-Saclay, Institut d'Optique Graduate School,CNRS, Laboratoire Charles Fabry, 91127 Palaiseau, France}

\author{Gianni Del Bimbo$^3$}
\thanks{These authors contributed equally to this work}
\author{Usman Ayub Sheikh$^3$}
\thanks{These authors contributed equally to this work}
\author{Maitree Shah$^4$}
\author{Luc Andrea$^5$}
\author{Faysal Ishtiaq$^3$}
\author{Andoni Duarte$^3$}
\author{Sam Mugel$^4$}
\author{Irene Cáceres$^3$}
\author{Michel Kurek$^5$}
\author{Roman Orús$^{3,6,7}$}
\affiliation{$^3$Multiverse Computing, Parque Cient\'ifico y Tecnol\'ogico de Gipuzkoa, Paseo de Miram\'on 170, 20014 San Sebasti\'an, Spain}
\affiliation{$^4$Centre for Social Innovation, 192 Spadina Ave, Suite 509, M5T 2C2 Toronto, Canada}
\affiliation{$^5$WIPSE Paris-Saclay Enterprises 7, rue de la Croix Martre 91120 Palaiseau, France}
\affiliation{$^6$Donostia International Physics Center, Paseo Manuel de Lardizabal 4, E-20018 San Sebasti\'an, Spain}
\affiliation{$^7$Ikerbasque Foundation for Science, Maria Diaz de Haro 3, E-48013 Bilbao, Spain}

\author{Achraf Seddik$^8$}
\author{Oumaima Hammami$^8$}
\author{Hacene Isselnane$^8$}
\author{Didier M'tamon$^8$}
\affiliation{$^8$Crédit Agricole Corporate and Investment Bank, 12 Place des États-Unis, 92545 Montrouge, France}

\date{\today}

\begin{abstract}
    Machine Learning models capable of handling the large datasets collected in the financial world can often become black boxes expensive to run. The quantum computing paradigm suggests new optimization techniques, that combined with classical algorithms, may deliver competitive, faster and more interpretable models.  
    In this work we propose a quantum-enhanced machine learning solution for the prediction of credit rating downgrades, also known as fallen-angels forecasting in the financial risk management field. We implement this solution on a neutral atom Quantum Processing Unit with up to 60 qubits on a real-life dataset. We report competitive performances against the state-of-the-art Random Forest benchmark whilst our model achieves better interpretability and comparable training times. We examine how to improve performance in the near-term validating our ideas with Tensor Networks-based numerical simulations.
\end{abstract}

\maketitle

\section*{Introduction}

In Finance, an interesting and relevant  problem consists in estimating the probability of debtors reimbursing their loans, which represents an essential quantitative problem for banks. Financial institutions generally attempt to estimate credit worthiness of debtors by sorting them in classes called credit ratings. These institutions can build their own credit rating model but can also rely on credit ratings provided by one or more of the three main rating agencies Fitch, Moody’s and Standard \& Poor’s (S\&P). Borrowers are generally grouped into two main categories according to their credit worthiness: investment grade borrowers with low credit risk, and sub-investment grade borrowers with higher credit risk. Should a borrower's rating downgrade from investment to sub-investment category, the borrower is referred to as a {\it fallen angel}.\\

The early anticipation of these fallen angels is a problem of utmost importance for financial institutions and one that has gathered significant attention from the machine learning (ML) community in the past years. Indeed, these institutions usually have access to large amount of data accumulated over several decades. The wide variety of features gathered can be fed to advanced ML models tasked with solving the following classification task: will a debtor have a high or low risk of becoming a fallen angel in the foreseeable future? Proposals of binary classification methods, or {\it classifiers}, targeting such tasks have been investigated and promising results were obtained using Random Forest and XGBoost \cite{credit_scoring_benchmark,xgboost_credit_scoring}. Due to their feature extraction flexibility, those tree-based ensemble methods turned out to be more suitable for similar credit risk modeling tasks \cite{ensemble_methods_credit,ensemble_methods_credit2}, compared to deep learning approaches \cite{gunnarsson2021deep,dastile2021making}. However, these methods quickly become computationally demanding as the numbers of decision trees grow. Furthermore denser and denser forests usually become black boxes in terms of interpretability, i.e. hard to understand by their users. \\

Quantum computing offers a new computational paradigm promising advances in computational efficiency for particular types of tasks. One of the most promising fields where quantum computing could be useful in practical industrial applications is the financial sector, with its wide range of hard computational problems \cite{Orus_2019}. Quantum and quantum-inspired approaches have already shown many promising applications in financial problems \cite{cacib_1, mugel2022dynamic, variational_clustering, Borle2020, Kyriienko2020, Paine2021, Kyriienko2021, Kyriienko2022}.  In particular, the nascent sub-field of Quantum Machine Learning harnesses well-known quantum mechanics phenomena like superposition and entanglement to enhance machine learning routines. This technology has drawn much attention in the past years with numerous proposed algorithms and a broad array of applications \cite{Schuld15, Biamonte17, Schuld19}. Indeed, many classification, clustering or regression tasks can be tackled by Quantum Neural Networks \cite{Mitarai2018, Kyriienko2020} or Quantum Kernel models \cite{Schuld2019, Henry2021, Paine2022,QEK_implem}. Much focus has also been placed on hybrid perspectives that combine classical components and current quantum hardware capabilities, like variational approaches \cite{variational_algos} or hybrid ensemble-type methods \cite{neven, neven1, denchev12, mott17, boyda17, dulny16}. \\

This paper investigates the capabilities of today's noisy intermediate-scale quantum hardware in conjunction with quantum-enhanced machine learning approaches for the detection of fallen angels. In this work, we propose a quantum-enhanced classifier inspired by the QBoost algorithm \cite{neven, neven1} with hardware implementation on a neutral-atom Quantum Processing Unit (QPU). We benchmark the proposed solution against a highly optimized Random Forest classifier. As we shall see, using the proposed quantum solution, we achieve a performance very similar to the benchmark with comparable training times and better interpretability, {\it i.e.} a simpler and smaller predictive model.  Furthermore, we provide a clear path on how we expect to beat the threshold set by the benchmarked Random Forest, providing evidence based on Tensor Networks simulations.\\

The paper is organized as follows: Section \ref{sec:classical solution} presents the application of the classical solution Random Forest on the classification task. Section \ref{sec:methodology} presents the proposed quantum-enhanced solution and methodology. Section \ref{sec:results} comments on the results and scaling of the devised quantum-enhanced classifier implemented on the proposed QPU, followed by conclusions and outlook.

\section{Classical Method for Fallen Angels Detection}
\label{sec:classical solution}

This section delves into the Random Forest model classically used in risk management to assess the creditworthiness of counterparts and predict future fallen angels. Random Forest is a well-known ensemble method based on bootstrap aggregation, also called {\it bagging}, applicable for regression and classification alike \cite{random_forest_1}. Training a random forest of $m$ trees on a $n$-size training set entails sampling with replacement the latter to generate $m$ new datasets with $n$ elements each. To ensure low correlation between the trees, each tree is trained on a different subset of randomly selected features. The trained classifiers are then collectively used to predict the class of unseen data through majority vote over $m$ decisions.

\subsection{Dataset} 
The dataset used in this study originates from public data over a period of $20$ years (2001-2020). It comprises of more than $90\,000$ instances characterized by around $150$ features, representing the historical evolution of credit ratings as well as numerous financial variables. Predictors include rating, financial and equity market variables and their trends calculated on a bi-annual, quarterly and five-year basis. The examples considered are based on over $2000$ companies from $10$ different industrial fields (e.g. energy, healthcare, utility) and $100$ sub-sectors (e.g. infrastructure, oil and gas exploration, mining), located in $70$ different countries. Each of the records is labeled as either a fallen-angel ({\it i.e.} critical downgrade; class 1/positive) or a non fallen-angel ({\it i.e.} stable or upgraded credit score; class 0/negative) based on standard credit rating scales. \\

The training set consists of around $65\,000$ examples from the 2001-2016 period. The testing set comprises of around $26\,000$ examples from the 2016-2020 period. The class distribution is highly unbalanced with only $9\%$ of fallen angels in the training set and $12\%$ in the test set. \\

Since our goal is to benchmark the performance of the proposed quantum framework against the classical Random Forest's solution over the same dataset, no further data processing or feature engineering was performed. To deal with the highly skewed distribution of classes as mentioned above, both random under-sampling of majority class and over-sampling of minority class were tested to balance the training set. \\

\subsection{Metrics}
\begin{figure}
    \centering
    \includegraphics[width=0.45\textwidth]{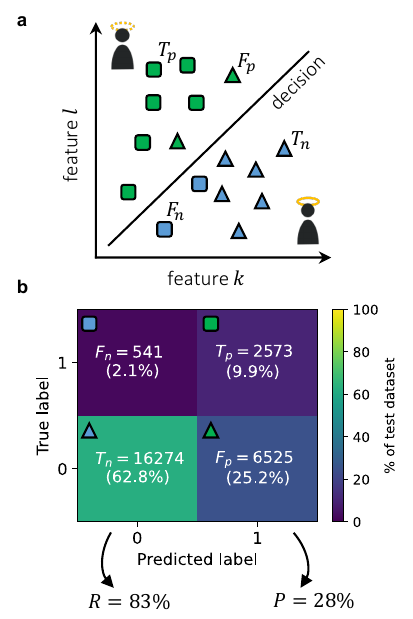}
    \caption{\textbf{a.} A binary classification problem deals with two classes (e.g. squares and triangles). The classifier is trained on the labeled training data and tasked to classify the instances, {\it i.e.} to find a decision boundary (black line) separating the test data. Considering that the test data comprising 2 features $k$ and $l$, the model either predicts correctly the class of an instance, thus increasing $T_{p/n}$ (green squares/blue triangles) or incorrectly, thus increasing $F_{p/n}$ (green triangles/blue squares).
   \textbf{b.} Classification performance of the classical solution on the test set shown as a confusion matrix. From the proportion of $T_p$ and $F_{p/n}$ obtained by the Random Forest model described in section \ref{ssec:classical_solution}, both recall $R$ and precision $P$ scores can be derived.}
   \label{fig:intro_to_classification}
\end{figure}

In order to assess the quality of the studied classification models, we settle on two metrics {\it i.e.} \emph{precision} and \emph{recall}, defined as:
\begin{equation}
    P=\frac{T_{p}}{T_{p}+F_{p}}\quad\text{and}\quad R=\frac{T_{p}}{T_{p}+F_{n}}.
\end{equation}
where $T_{p/n}$ represents the number of true positives/negatives ({\it i.e.} accurate prediction of the model) and $F_{p/n}$ represents the number of false positives/negatives ({\it i.e.} inaccurate prediction of the model) as illustrated in Fig.\,\ref{fig:intro_to_classification}\textbf{a}. The precision $P$ is the ability of a classifier to not mistake a negative sample for a positive one; it thus represents the quality of a positive prediction made by the model. Similarly, the recall $R$ can be understood as the ability of the model to find all the positive samples.\\

Our primary goal consists in increasing the precision of predicting fallen angels while keeping the recall over $R=80\%$. Accommodating this criterion requires tuning the decision threshold, a parameter that governs conversion of class membership probabilities to the corresponding hard predictions (e.g. $0$ or $1$).  This threshold is usually set at $0.5$ for normalized predicted probabilities. \\

\begin{figure}
    \centering
    \includegraphics[width=0.45\textwidth]{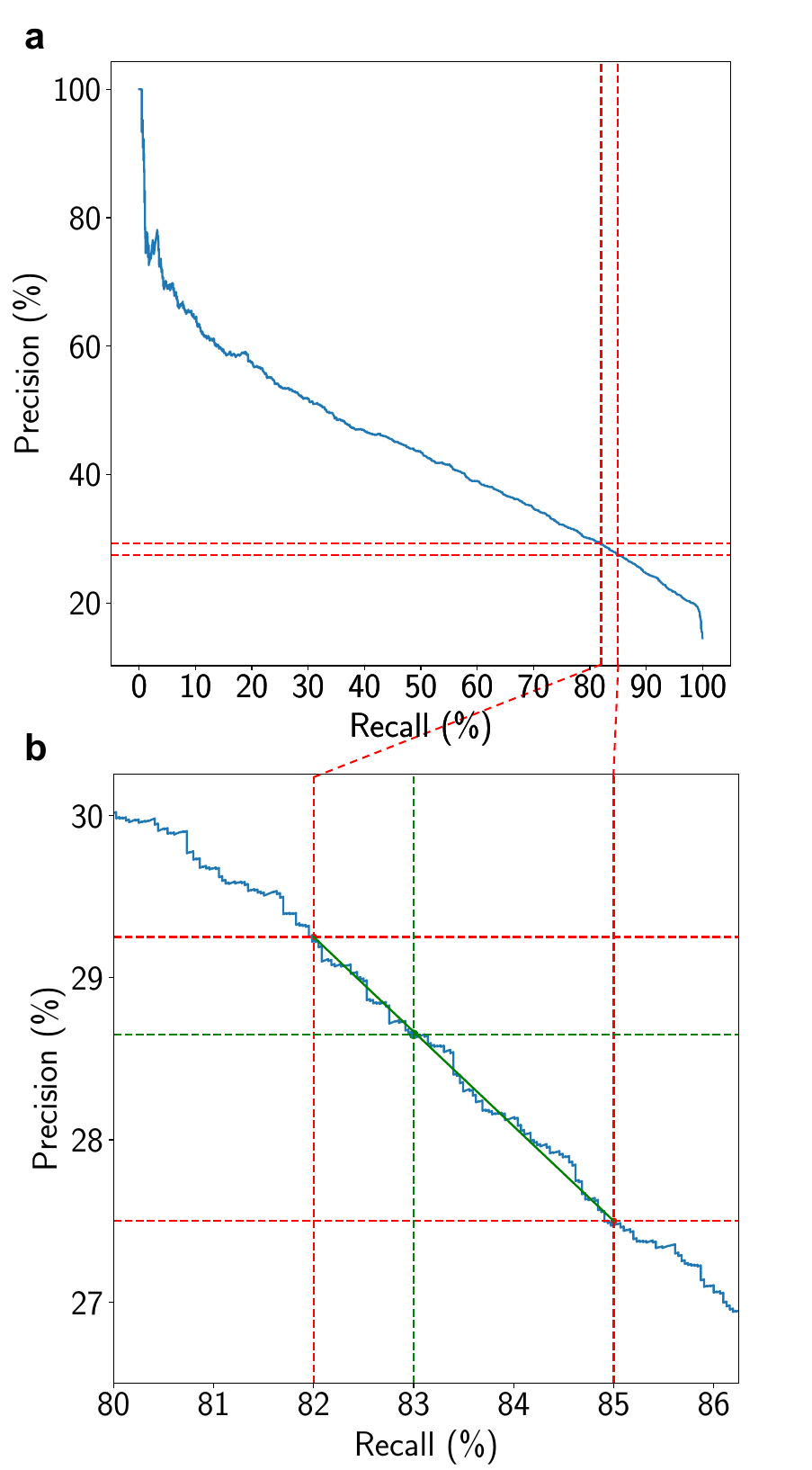}
    \caption{\textbf{a} Precision-Recall curve (blue) calculated using different decision thresholds for the predictions. A close-up of the region of interest is shown in the following (dashed-red). \textbf{b} Precision value corresponding to the recall of 83\% obtained through linear interpolation (green) between the neighboring points $R=82\%$ and $85\%$ on the PR curve.}
    \label{fig:pr_curve}
\end{figure}

Here, the optimal probability threshold value is determined using a Precision-Recall (PR) curve shown in Fig.\,\ref{fig:pr_curve}\textbf{a}. Specifically, $P$ and $R$ are computed at several decision thresholds. A linear interpolation between neighboring points of the PR curve (see Fig.\,\ref{fig:pr_curve}\textbf{b}) enables to determine the precision value corresponding to  $R=83\%$.

\subsection{Benchmark Baseline}
\label{ssec:classical_solution}
Training the Random Forest model and optimizing its hyperparameters through random search, lasts more than 3 hours on a classical computer. The 1200 decision trees of the obtained model enable to achieve $R=83\%$ and $P=28\%$, as showcased in Fig.\,\ref{fig:intro_to_classification}\textbf{b}. 

This result, far from being optimal, especially in terms of precision, is due to several factors, representative of the complexity of the problem:
\begin{enumerate}
    \item  The dataset is highly unbalanced, which is a notoriously hard problem for classical machine learning models.
    \item Processing a significant amount of features can be resource-consuming and it remains impossible to exhaustively search the space of solutions at too large sizes. Therefore, the classical method uses a suboptimal shortcut to select relevant features.
    \item Finding the optimal weight for each predictor is an exponentially complex optimization problem as the number of predictors increases. Hence, the Random Forest model uses majority voting for classification, which is quite restrictive in terms of performance.
\end{enumerate}
We address the above-mentioned points with a proposed quantum-enhanced machine learning approach, taking advantage of quantum combinatorial optimization to efficiently explore the space of solutions. 

\section{Quantum-enhanced Classifier}
\label{sec:methodology}
\subsection{QBoost}
\label{ssec:qboost}
First proposed by Neven et al. \cite{neven}, the QBoost algorithm is an ensemble model comprising of a set of weak, i.e. simple low-depth, Decision Tree (DT) classifiers, also called learners, optimally combined to build a strong classifier. \\

The workflow of the algorithm starts from a boosting procedure, based on the standard Adaboost algorithm \cite{freud97,Wang_2012}. A set of $N$ weak learners $h_{i=1\hdots N}$ is classically trained in a sequential fashion on the training set. Initially, the \textit{first} weak learner is trained such that all the data points are weighted uniformly, using the constant distribution
$D_{i=1}(s)=\frac{1}{S}$, with $S$ the size of the dataset.  Each weak learner $h_{i}$ is then iteratively trained on the same training set where the data points are however weighted differently based on an updated distribution $D_{i}(s)$. This latter distribution is re-computed after the training of each weak learner. More precisely, it depends on the quantity $\varepsilon_{i}$ that considers the misclassified points by the previous weak learner:
\begin{equation}
    \varepsilon_{i}=\sum_{s}{\mathbbm{1}\left[h_{i}(\vec{x}_s)\neq y_{s}\right]}D_{i}(s).
\end{equation}
From $\varepsilon_{i}$, one computes the quantity
\begin{equation}
    \alpha_{i}=\frac{1}{2}\ln\left(\frac{1-\varepsilon_{i}}{\varepsilon_{i}}\right),
\end{equation}
which enters the exponential coefficient to update the data distribution as
\begin{equation}
    D_{i+1}(s)=\frac{1}{Z_i}D_{i}(s)\cdot e^{-\alpha_{i}y_{s}h_{i}(\vec{x}_s)}.
\end{equation}
Here, we introduce the normalization factor $Z_{i}$ such that $D_{i+1}$ is a probability distribution. \\

After the entire ensemble of weak learners $h_{i=1\hdots N}$ has been trained, a strong classifier $C$ is built by combining the weak learners. The optimal combination is obtained through the optimization of  binary weights $w\in\mathbb{B}^N$ that minimize the following cost function 
\begin{equation}
    \mathcal{H}(w) = \sum_{s}\left(\frac{1}{N}\sum_{i}^{N}w_{i}h_{i}(\vec{x}_{s})-y_{s}\right)^{2}+\lambda\Vert w\Vert_{0},
\end{equation}
where $w_i$ is the $i$-th binary weight, $h_{i}(\vec{x_s})\in[-1,1]$ is the prediction of the $i$-th weak learner for the data point $\vec{x}_{s}$, and $y_{s}\in[-1,1]$ the classification labels.  A regularization term parameterized by $\lambda$ helps to favor better generalization of the model on new data by penalizing too complex ensembles with many weak learners. As the number of learners increases, the space of possible binary weigths expands exponentially. Minimizing $\mathcal{H}$ is in fact an NP-hard problem. \\

Expanding the squared term in the above equation and dropping the constant terms, which are irrelevant to the minimization problem, we can  reformulate the cost function as 
\begin{equation}
    \begin{split}
        \mathcal{H}(w) = & \sum_{i,j}^{N}w_{i}w_{j}{\rm Corr}(h_{i}, h_{j})\\
         & +\sum_{i}^{N}w_{i}(\lambda-2{\rm Corr}(h_{i},y)),
    \end{split}
\label{eq:cost_function}
\end{equation}
with ${\rm Corr}(h_{i},h_{j})=\sum_{s}h_{i}(\vec{x}_{s})h_{j}(\vec{x}_{s})$ and ${\rm Corr}(h_{i}, y)=\sum_{s}h_{i}(\vec{x}_{s})y_{s}$. On the one hand, the weak classifiers whose outputs correlate well with the labels cause the second term to be lowered via ${\rm Corr}(h_{i},y)$. On the other hand, via the quadratic part  ${\rm Corr}(h_{i}, h_{j})$ describing the correlations between the weak classifiers, pairs of strongly correlated classifiers increase the value of the cost function, thereby increasing the chance for one of them to be switched off. This is in line with the general paradigm of ensemble methods for promoting a diversification of the ensemble in order to improve the model generalization on unseen data.  \\

Once the optimization of Eq.(\ref{eq:cost_function}) is performed (see section \ref{ssec:qubo_optimization}), the strong classifier $C$ can be built using $w^{opt}$, the weights minimising $\mathcal{H}$. Given a new data point $\vec{x}$, we infer a classification prediction by:
\begin{equation}
    C(\vec{x})={\rm sign}\left(\frac{1}{N}\sum_{i}^{N}w_{i}^{opt}h_{i}(\vec{x})-T\right)
\end{equation}
where $T$ is an optimal threshold that enhances results as proposed in \cite{neven, neven1} and computed as a post-processing step 
\begin{equation}
    T=\left(\frac{1}{S}\sum_{s}^{S}\frac{1}{N}\sum_{i}^{N}w_{i}^{opt}h_{i}(\vec{x}_s)\right).
\end{equation}

\subsection{QBoost-inspired Quantum Classifier}
\label{ssec:proposed_classifier}

An important challenge in designing a successful ensemble is to ensure that the base learners are highly diverse, {\it i.e.}, that their predictions do not correlate too much with each other. The initial idea of QBoost \cite{neven} was to accomplish this by using weak learners of the same type, specifically Decision Trees (DT) classifiers and train them sequentially using the boosting procedure. Another way is to use different types of base learners \cite{van2018online}, creating an heterogeneous ensemble with a mix of different learners including, e.g., DT, Logistic Regression (LR), k-Nearest Neighbors (kNN), and Gaussian Naive Bayes (GNB) \cite{bishop2006pattern}. Having inherently different mathematical foundations, these learners can exhibit significantly different views of the data landscape.  \\

\begin{figure}
    \centering
    \includegraphics[width=\linewidth]{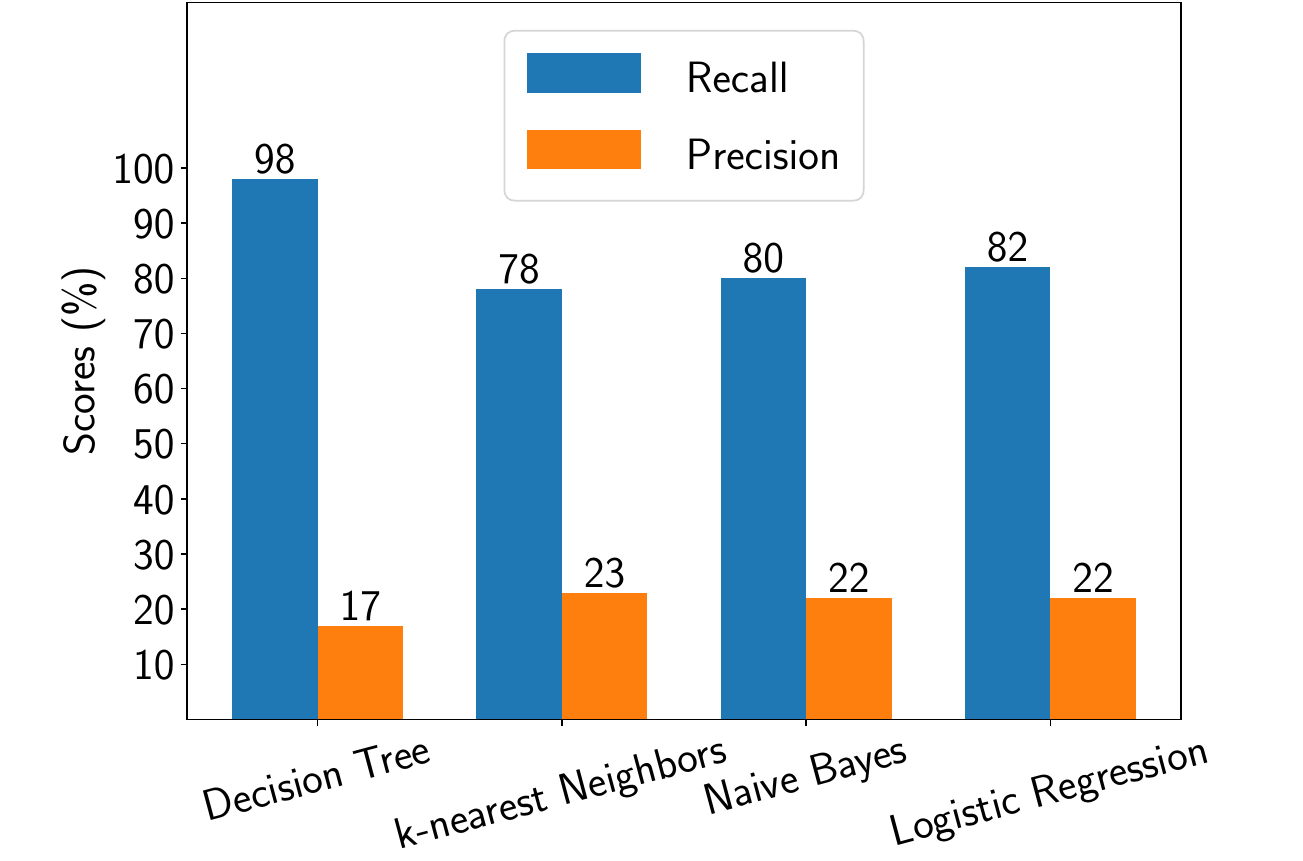}
    \caption{Comparison of QBoost performances, {\it i.e.} recall (blue) and precision (orange), with different base learners including decision trees, k-nearest neighbors, gaussian naive bayes and logistic regression.}
    \label{fig:weak_learners}
\end{figure}

For this specific problem, we find that a classifier based on a heterogeneous ensemble comprising different types of learners can lead to better generalization performance than the plain-vanilla model with DT only. The choice of type and mixing of such an heterogeneous ensemble is motivated by comparing results from extensive simulations, both with one type of learner as well as with combinations thereof. Each of these models are trained on the under-sampled version of the training set and the corresponding prediction performance are obtained on a separate cross-validation set, using Precision and Recall. As displayed in Fig.\,\ref{fig:weak_learners}, DTs perform better in recall while kNNs perform better in precision. Heuristically, combining these two types of base learners results in the actual best performing model over any other tested combinations.\\

Furthermore, in order to take advantage of the historical structure of the data, where multiple historical data points for the same companies are available, we propose to train each of the learners of the heterogeneous ensemble on different historical periods of the dataset. Specifically, using dates features in the dataset, the raw training set is split into subsets and then subgroups of learners are trained on them. This ensemble-training procedure based on subsampling is expected to further diversify the ensemble, where the weak learners are trained independently on the different economic recession and expansion periods underlying the training dataset. Additionally, it reduces the training time significantly as each subgroup of learners is trained on a subset of data. The learners trained in this way can potentially capture different views of the data, resulting in a better diversification of the ensemble.\\

Here we propose this approach with two variations:
\begin{enumerate}
    \item \textit{Boosting}. Following \cite{neven}, we train each of the ensembles on the different subsamples of data with the sequential boosting procedure described in section \ref{ssec:qboost}. Generally, the learners can exhibit negative correlations among each other.
    \item \textit{Subsampling}. We train each of the ensembles without sequential boosting, relying only on subsampling for diversification. Generally, the learners exhibit only positive correlations among each other.
\end{enumerate}

\subsection{Optimization of the Ensemble via QUBO Solving}
\label{ssec:qubo_optimization}
As explained in section \ref{ssec:qboost}, the weak learners obtained during the ensemble training are then optimally combined to
form a stronger classifier. Finding the best binary weights $w$ for this combination amounts to the minimization of the cost function given in Eq.(\ref{eq:cost_function}). Since the weights $w_{i}$ are binary variables, that is, $w_{i}^{2}=w_{i}$, we reformulate $\mathcal{H}$ as $\mathcal{H}_Q$:
\begin{equation}
    \mathcal{H}_Q(w) =w^TQw= \sum_{i,j}^N Q_{ij}w_{i}w_{j},
\label{eq:qubo_formula}
\end{equation}
where
\begin{equation}
    Q_{ij}=\left\{
    \begin{array}{ll}
        {\rm Corr}(h_{i},h_{j}) & \mbox{ if } i\neq j \\
        \frac{S}{N^{2}}+\lambda-2{\rm Corr}(h_{i},y) & \mbox{ else.}
    \end{array}
    \right.
\end{equation}
This second formulation is written in the form of a \textit{Quadratic Unconstrained Binary Optimization} (QUBO) problem \cite{QUBO}. Solving a QUBO problem amounts to finding the minimum of a quadratic polynomial of bit variables, {\it i.e.}, 
the optimal bitstring minimizing the cost function $\mathcal{H}_Q$, with $Q\in M_N(\mathbb{R})$, the symmetric matrix encompassing the correlations between the weights to optimize. \\

As the number of learners grows, the classical optimization of the weights becomes exponentially hard, thus opening the door to potentially more efficient quantum methods.
For instance, what if we encode the binary variables into qubit states on a quantum computer, traversing the search space as a Hilbert space? A common misconception is that quantum computers could solve in polynomial time NP-complete problems, but this has not been proven and in fact the consensus is that it would be extremely unlikely. However, there is mounting evidence \cite{QAOA} that they could better \textit{approximate}  `sufficiently good’ (as defined in Section \ref{ssec:results_qubo_solving}) solutions in a short(er) time compared to classical computers in some cases. This expectation partly stems from the fact that quantum computers may offer shortcuts through the optimization landscape inaccessible to traditional classical simulated annealing methods \cite{Crosson2016}. In our case, if one can produce a state such that the probability amplitudes peak in low-cost bitstrings, sampling from it becomes an efficient way of optimizing the weights. Such a quantum state may be potentially highly entangled, and cannot be efficiently stored by classical means. Quantum computers based on neutral atoms offer unprecedented scalability, up to hundreds of atoms \cite{324}, as well as a global addressing scheme (analog mode), allowing a large set of qubits to be easily entangled. In contrast, quantum circuit-based calculations become quite greedy regarding the number of gates required to achieve such levels of complexity. We focus here on solving QUBO problems using an analog neutral atom setup. 

\subsection{Information Processing  on a Neutral Atom platform}\label{ssec:qubo_qpu}
In analog neutral atom quantum processing devices \cite{Henriet_2020}, each atom is considered with reasonable approximation as a simpler system described by only two of its electronic states. Each atom can thus be used as a qubit with basis states $\ket{0}$ and $\ket{1}$, being respectively a low-energy {\it ground} state and a highly excited Rydberg state \cite{vsibalic2018rydberg}. The evolution of the qubits state can be parameterized by time-dependent control fields, $\Omega(t)$ and $\delta(t)$. These parameters are ultimately related to the physical properties of lasers acting on the atoms. Moreover, the quantum state of atom $i$ can significantly alter the state of atom $j$, depending on their pair distance $r_{ij}$. Indeed, the excitation of $i$ to a Rydberg state $\ket{1}$ shifts the energies of the corresponding Rydberg states of nearby $j$ by an amount $U(r_{ij})$. The latter quantity can be considered impactful compared to the action of the control fields when $r_{ij}\leq r_b$, with $r_b$ the \emph{blockade radius}. This blockade effect \cite{lukin2001dipole} constitutes the building block of the entangling process in neutral atom platforms.\\

When a laser pulse sequence acts on an entire array of $N$ atoms, located at positions $\mathbf{r}$, the time evolution of the quantum state $\ket{\psi}$ can be expressed by the Schrödinger equation $i\hbar\frac{\partial }{\partial t} \ket{\psi} = \hat H(t) \ket{\psi}$, where $\hat H(t)$ is the system Hamiltonian. Neutral atom devices are capable of implementing the so-called \emph{Ising} Hamiltonian, consisting of a time-dependent control part as well as a position-dependent interaction part : 
\begin{equation}
\begin{split}
    \hat{H}(t)=&\,\hat{H}_{ctrl}(t)+\hat{H}_{int}(\mathbf{r})\\
    =&\,\hbar\sum_{i=1}^N\left(\frac{\Omega(t)}{2} \hat\sigma_i^x - \delta(t) \hat n_i\right)+\frac{1}{2}\sum_{i\neq j}U_{ij}\hat n_i \hat n_j,
\end{split}
\label{eq:rydberg_ham}
\end{equation}
where $\hat \sigma_i^{\alpha}$ are the Pauli matrices applied on the $i^{th}$ qubit\footnote{The Pauli matrices are:
$$
\hat \sigma^x = \begin{pmatrix}
    0&1\\
    1&0
  \end{pmatrix}, \, \\
\hat \sigma^y =
  \begin{pmatrix}
    0& -i \\
    i&0
  \end{pmatrix}, \,\\
\hat \sigma^z =
  \begin{pmatrix}
    1&0\\
    0&-1
  \end{pmatrix}
$$
The application of a Pauli matrix on the $i^{th}$ site of the quantum system is represented by a tensor (or Kronecker) product of matrices: $\sigma_i^\alpha = \mathbb I \otimes \cdots \otimes \sigma^\alpha \otimes \cdots \mathbb I$, for $\alpha=x,y,z$. For instance,  $\hat\sigma_1^x\ket{00}=\ket{10}$. A similar construction is used for composite operators like $\hat n_i \hat n_j$.}, $\hat n_i =(1+\hat \sigma_i^z)/2$ is the number of Rydberg excitations (with eigenvalues $0$ or $1$) on site $i$, and $U_{ij}=U(r_{ij})>0$ represents the distance-dependent interaction between qubit $i$ and $j$. The state of the system is initialized to $\ket{0\hdots 0}$. Once the pulse sequence drives the system towards its final state $\ket{\psi}=\sum_{w\in\mathbb{B}^N}a_w\ket{w}$, a global measurement is performed through fluorescence imaging: the system is projected to a basis state $\ket{w}$ with probability $\abs{a_w}^2$. The obtained picture reveals which atoms were measured in $\ket{0}$ (bright spot) and which were in $\ket{1}$ (dark spot), see Fig.\,\ref{fig:RGS_process}. Repeating the cycle (loading atoms, applying a pulse sequence and measuring the register) multiple times constructs a probability distribution that approximates $\abs{a_w}^2$ for $w\in\mathbb{B}^N$, allowing to get an estimator of $\ket{\psi}$ and use it as a resource for higher-level algorithms.  

\subsection{Quantum Algorithms for QUBO Problems}

Proposals to solve combinatorial problems, like QUBOs, using neutral atom quantum processors abound \cite{Ebadi_2022,pichler_MIS,parity}, for example using a Quantum Adiabatic Algorithm (QAA)\,\cite{Farhi_2001} or Quantum Approximate Optimization Algorithm (QAOA)\,\cite{QAOA}. One crucial ingredient of these proposals is the ability to implement a custom cost Hamiltonian $\hat{H}_Q$ on the quantum processor which should be closely related to the cost function $\mathcal{H}_Q$. When this Hamiltonian is generated exactly, the mentioned iterative methods can ensure that after $k$ iterations the evolution of a quantum system subjected to $\hat H^{(k)}_{ctrl}+\hat H_Q$  will tend to produce low energy states $\ket{w}$, {\it i.e.} solutions  with low values $\mathcal{H}_Q(w)$. There are ways to compute the evolution over $\hat{H}_Q$ using circuit-based quantum computers\,\cite{QAOA}, or special-purpose superconducting processors like D-wave machines \cite{Zbinden2020-ol}. \\

For analog neutral atom technology, innately replicating the cost Hamiltonian with $\hat H_{int}$ would require to satisfy $\{U_{ij}=Q_{ij}\| 1 \leq i\neq j\leq N\}$, and thus to find specific coordinates of the atoms respecting all these constraints (their number increases quadratically with the number of variables used to represent $N$ atoms in the plane). As a consequence, only part of the constraints can be fulfilled by a given embedding, resulting in only an approximation of $\hat H_Q$ in a general case. Therefore, the quality of the solutions sampled from the final quantum state will be limited under a straightforward implementation of QAA and QAOA. In addition, the optimization of variational quantum algorithms \cite{variational_algos} usually requires diagnosing the expressibility and trainability of several circuits (or pulse sequences at the hardware level) in order to trust that low energy states are being constructed. Moreover, at each iteration, obtaining a statistical resolution of the energy of the prepared state with precision $\varepsilon$, one requires $\mathcal O (1/\varepsilon^{2})$ samples. Given the currently low repetition rate of Quantum Processing Units (QPUs) based on neutral atom technology (of the order of $1-5$ Hz), the implementation of such approaches requires several tens of hours of operation on robust hardware. With current technology, it is therefore crucial to employ methods involving only a small budget of cycles and that can quickly provide significant solutions to the QUBO problem. 

\subsection{Random Graph Sampling}
\label{ssec:RGS}

\begin{figure*}
    \centering
    \includegraphics[width=0.95\textwidth]{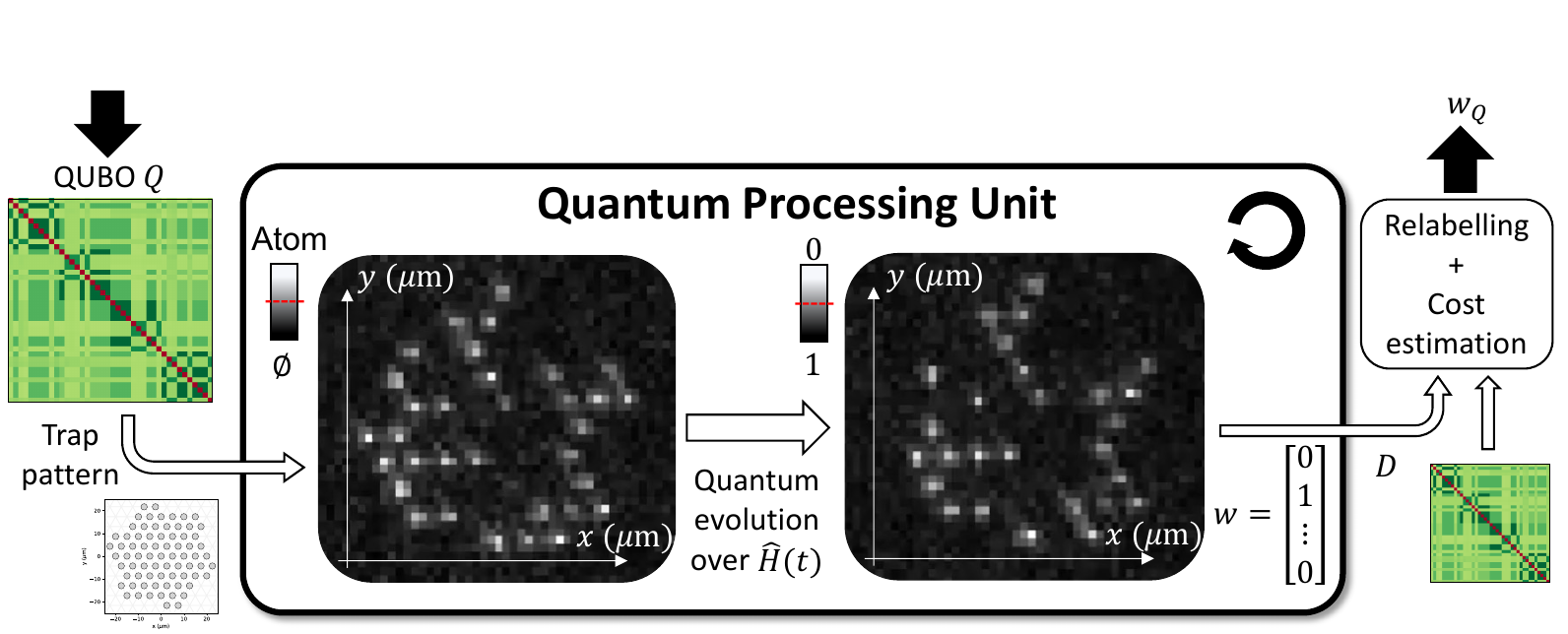}
    \caption{Random Graph Sampling pipeline for solving a QUBO $Q$ on a neutral atom based QPU. First, a QUBO, here with negative weights on the diagonal (red) and positive weights outside (green scale), is taken as input. From this QUBO, a trap pattern is devised and sent to the QPU, as well as the wanted number of repetitions and the pulse sequence used. At the beginning of each cycle, a first fluorescence picture enables to identify which traps were filled by an atom (bright spot). The system evolves according to $\hat H(t)$ and a final picture is taken to measure the collapsed state of the system, outputting a bitstring $w$. Using $Q$, we select $w_Q$ among $D$, the distributions of bitstrings obtained from repeating this process several times.}
    \label{fig:RGS_process}
\end{figure*}

Stepping away from the variational paradigm of QAOA and QAA, we devised a sampling algorithm that exploits the stochastic loading probability of neutral atom QPU in order to probe efficiently the solution space of a QUBO. This algorithm is faster to implement as it does not require iterative processes such as closed loop communication between a classical optimizer and the quantum hardware. We propose the Random Graph Sampling (RGS) method which builds up on the randomness of the atomic loading process. This procedure allows us to sample solutions of QUBOs of sizes up to $60$. \\

In neutral atom QPU, atoms are spatially arranged by combining the trapping capacity of optical tweezers with the programmability of a Spatial Light Modulator (SLM). By the means of those two devices, atoms can be individually trapped in arbitrary geometries. Once an atomic cloud has been loaded, each of the $N_t$ traps is randomly filled with success probability $p$ according to a binomial law $\mathcal{B}(N_t,p)$. Thus, one must set up around $N_t=N/p$ traps to maximize the probability of trapping $N$ atoms. A rearrangement algorithm then moves atoms one at a time to the wanted positions using a moving tweezer. The excess atoms are released mid-stroke to end up with a register of $N$ correctly positioned atoms. However, by skipping the rearrangement step, we obtain samples from an essentially random sub configuration of the underlying pattern of traps. For $N_t=2N$, the number of possible configuration of size $N$ scales as $\binom{2N}{N}\sim 4^N/\sqrt{N}$, offering a large variety of $\hat H_{int}$ for a devised pattern. In order to produce the quantum distributions from which we sample the QUBO solutions, we repeatedly apply a parameterized sequence with constant pulses to the atoms. The latter evolve under $\hat H(t)$ according to their interactions, which are set by the atom positions at each cycle.\\

The QUBO to solve, $Q$, first acts as a resource to design the trap pattern ($N_t$, shape, spacing) sent to the QPU (see Fig.\,\ref{fig:RGS_process}). Once a chosen budget of samples has been acquired on the QPU, we are left with a bitstring distribution $D$. Using again the QUBO, we apply a relabeling procedure (described in Appendix \ref{app:RGS_details_opti}) to each bitstring according to both $Q$ and the related atom positions. This optimization procedure is designed to scale only linearly with $N$ and is tasked to search for a way of labeling the atoms from $1$ to $N$ which minimizes for each repetition the difference between $\hat H_Q$ and $\hat H_{int}$. Finally, we compute the bitstring corresponding to the optimized weights for the ensemble of learners considered.\\

While RGS offers no theoretical guarantee of sampling a global minimum of the cost function, we can still expect to output bitstrings with low function value. We can compare RGS performances to several state-of-the-art quantum-inspired methods. Quantum-inspired algorithms are run on classical devices and allow with some approximation a fast emulation of the quantum phenomena happening in a QPU. Two numerical methods are introduced to benchmark QUBO solving: a naive analog QAOA with $3$ pulse durations as optimizable parameters (see Appendix \ref{app:RGS_details_opti}) and Simulated Annealing (SA) \cite{SA,Das_2008}. The methods are always compared for similar budget of cycle repetitions and we can assess the quality of the bitstring distribution or the scalability of each approach. We also propose in the following a numerical Tensor Networks-based algorithm to handle large QUBOs without any restriction on their structure. 

\subsection{Tensor Network Optimization}
\label{ssec:tn}
Tensor Networks (TN) are a mathematical description for representing quantum-many body states based on their entanglement amount and structure \cite{orus2014practical,orus2019tensor}. They are used to decompose highly correlated data structures, {\it i.e.}, high-dimensional vectors and operators, in terms of more fundamental structures and are especially efficient in classically simulating complex quantum systems. \\

To be more specific, one can consider an $N$-qubits system and its wave function naively described by its $O(2^N)$ coefficients $a_w$ in the computational basis. Formally, these coefficients can be represented by a tensor with $N$ indices, each of them having two possible values ($0$/$1$), which quickly become costly to store and process with increasing $N$. However, we can replace this huge tensor by a network of interconnected tensors with less coefficients, defining a TN. Each subsystem corresponds in practice to the Hilbert space of one qubit. By construction, the TN depends on $O(poly(N))$ parameters only, assuming that the rank of the interconnecting indices is upper-bounded by a parameter, called {\it bond dimension}. Because of polynomial scaling, TN constitute useful tools for emulating quantum computing, and many of the current state of the state-of-the-art simulators of quantum computers are precisely based on them.\\

In addition, TNs have also proven to be a natural tool for solving both classical and quantum optimization problems \cite{mugel2022dynamic}. They have been used as an ansatz to approximate low-energy eigenstates of Hamiltonians. In our case, which involves a classical cost function, we propose an optimization algorithm based on Time-Evolving Block Decimation (TEBD) \cite{vidal2003efficient}. In this approach, we simulate an imaginary-time evolution driven by the classical Hamiltonian $\hat H_Q$, and simulate the state of the evolution at every step by a particular type of tensor network called Matrix Product State (MPS). By using this approach, the algorithm reaches an optimum final configuration of the bitstring minimizing the cost function. The optimal configuration of the ensemble thus achieved was used to make predictions on the unseen test set. 

\section{Results}
\label{sec:results}

In this section, we present the classification results based on the RGS quantum optimization procedure performed on QPU for QUBOs of sizes up to $60$ qubits. 
%We find that improving the proposed best bitstring requires a comparable number of repetition steps as that of standard quantum-inspired tools for QUBO solving, like Simulated Annealing. We also report that the implementation of the quantum classifier with subsampling on the QPU exhibits competitive results  with the Random Forest benchmark. 
In addition, we present the performance of the boosting variation of the quantum classifier, already capable of beating the benchmark by reaching a precision of $29\%$ for $90$ qubits/learners, based on Tensor Networks methods.
%These last results are obtained using the quantum-inspired optimization algorithm based on Tensor Networks (see Section \ref{ssec:tn}).  We also highlight the relevant limitations of current technology and the upgrades needed to be able to replicate the boosting approach results on the QPU. 

\subsection{QPU Optimizer Results}
\label{ssec:results_qubo_solving}

\begin{figure}
    \centering    
    \includegraphics[width=\linewidth]{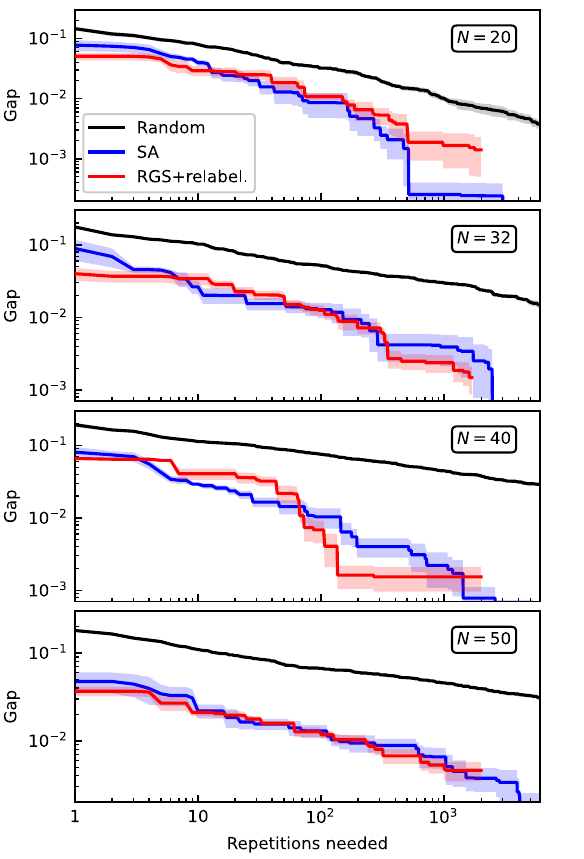}
    \caption{Gap convergence obtained with classical uniform sampling (black), Simulated Annealing sampling (blue) and RGS sampling with optimized relabeling (red) for increasing size of QUBOs. The best gap found after some cycle repetitions is averaged over sets of $5$ QUBOs (plain line). \textbf{b.} Scaling analysis of the number of repetitions needed to reach a gap below a threshold of $1\%$ with respect to problem size. The results obtained by the three mentioned methods at sizes $N= 12$, $20$, $32$, $40$, $50$ and $60$ (dots) are fitted either exponentially or polynomially (line) depending on the best match.}
    \label{fig:results-sizes}
\end{figure}

We experimentally implement the RGS method described in section \ref{ssec:RGS} to solve $6$ sets of QUBOs ranging in sizes from $N=12$ to $60$ qubits. QUBOs of one set being produced by repeatedly applying the subsampling approach on the same dataset, they only exhibit minor discrepancies between their structure and range of values. Therefore, we can, within reasonable approximation and for faster implementation on the quantum hardware, only use one trap pattern per set. In addition, we can reuse statistics acquired at large sizes and extract from them distributions of bitstring of smaller size as explained in Appendix \ref{app:RGS_details_clustering}. In essence, by neglecting the interaction between pairs of atoms separated by more than one site, a $N$-atom regular array can be divided into smaller clusters with similar regular shape and isolated from each other. \\

\begin{figure}
    \centering    
    \includegraphics[width=\linewidth]{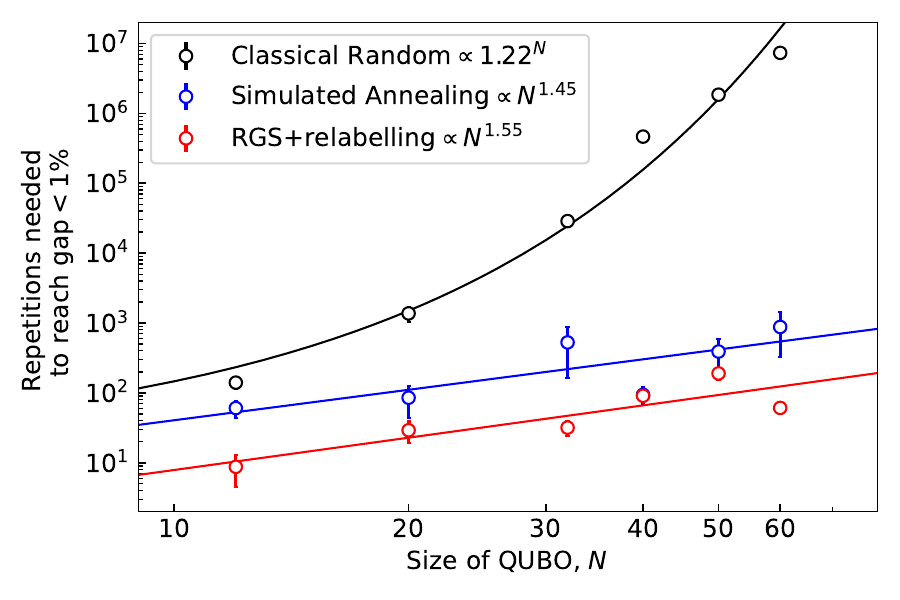}
    \caption{Scaling analysis of the number of repetitions needed to reach a gap below a threshold of $1\%$ with respect to problem size. The results obtained by the three mentioned methods at sizes $N= 12$, $20$, $32$, $40$, $50$ and $60$ (dots) are fitted either exponentially or polynomially (line) depending on the best match.}
    \label{fig:scaling-sizes}
\end{figure}

Considering a loading probability $p=0.55$, we design a triangular pattern with $N_t=40/0.55\approx73$ traps for QUBOs of size $40$ (see Fig.\,\ref{fig:RGS_process}) and similarly with $91$ traps for QUBOs of size $50$. While this choice is motivated by both available trapping laser power and maximization of number of samples at sizes $40$ and $50$, it restricts the number of statistics gathered at size $60$. Results at this size are thus subject to large uncertainty bars. The spacing of the regular pattern and thus the atomic interaction in the array is chosen in combination with the maximum value of $\Omega(t)$ reached during the pulse sequence. Having Hamiltonian terms $\Omega$ and $U$ of comparable magnitude in $\hat H(t)$ enables to explore the strongly interacting regime.\\

Finally, we introduce the \emph{gap} of a bitstring $w$ defined as:
\begin{equation}
\label{eq:gap}
    \text{gap} = \abs{\frac{\mathcal{H}_Q(w)-\mathcal{H}_Q(w_Q^0)}{\mathcal{H}_Q(w_Q^0)}}
\end{equation} where $w_Q^0$ is the best solution returned by a state-of-the-art SA algorithm given a large amount of repetitions ($200\,000$ here). This solution $w_Q^0$ is not guaranteed to be the best possible, but acts as such for benchmark purposes. Reaching a gap of $0$ amounts to having found the best solution prodived by the benchmark. Note that for small sizes of $N$, it is possible to use an exhaustive search as a benchmark and $w_Q^0$ is in that case the theoretical best solution. Finding a bitstring with a gap below $1\%$ for instance, means finding a solution with a cost close to $1\%$ of the optimal one, which, in many operational problems such as our case study, is often sufficient. We check that for the various sizes considered, the difference between selecting a $1\%$ solution and the optimal one, {\it i.e.} with a gap of $0$) is reflected in the classification model with variation of precision $P$ smaller than the standard deviation obtained on the QUBO set. We thus consider as a good enough solution a bitstring with gap below $1\%$.\\

The results obtained by RGS with relabeling are showcased both in terms of convergence to low cost value $\mathcal{H}_Q(w)$ solutions (see Fig.\,\ref{fig:results-sizes}) and scalability of the method with respect to the complexity of the problem, {\it i.e.} the QUBO size (see Fig.\,\ref{fig:scaling-sizes}). The classical random method consisting in uniformly sampling with replacement bitstrings from $\mathbb{B}^N$, it scales exponentially with $N$. In contrast, the RGS algorithm shows better performances, already finding solutions with a gap smaller than $10\%$ after only a few repetitions. Looking at the number of repetitions needed to go below $1\%$ with respect to $N$, a log-log linear fit returns a scaling in $0.2\times N^{1.55}$. Since the QPU run-time scales linearly with the number of cycles, the quantum optimization duration is also expected to scale polynomially. Comparing RGS to the SA algorithm, we observe better performance of the latter at small sizes but more and more comparable performance at increasing sizes. In the case of $N=40$, this specific implementation of RGS finds on average a gap below $0.2\%$ after $150$ repetitions while SA needs around $4$ times more cycles. For $N=60$, the mean gap achieved after hundreds of cycle is around $1.5\%$.  Overall, RGS with relabeling applied to QUBOs produced by the subsampling-based classifier exhibits similar behavior with state-of-the-art SA algorithm. 

\subsection{QPU Classification Results}
\label{ssec:qpu_classification_results}
\begin{figure*}[]
    \centering
    \includegraphics[width=0.9\textwidth]{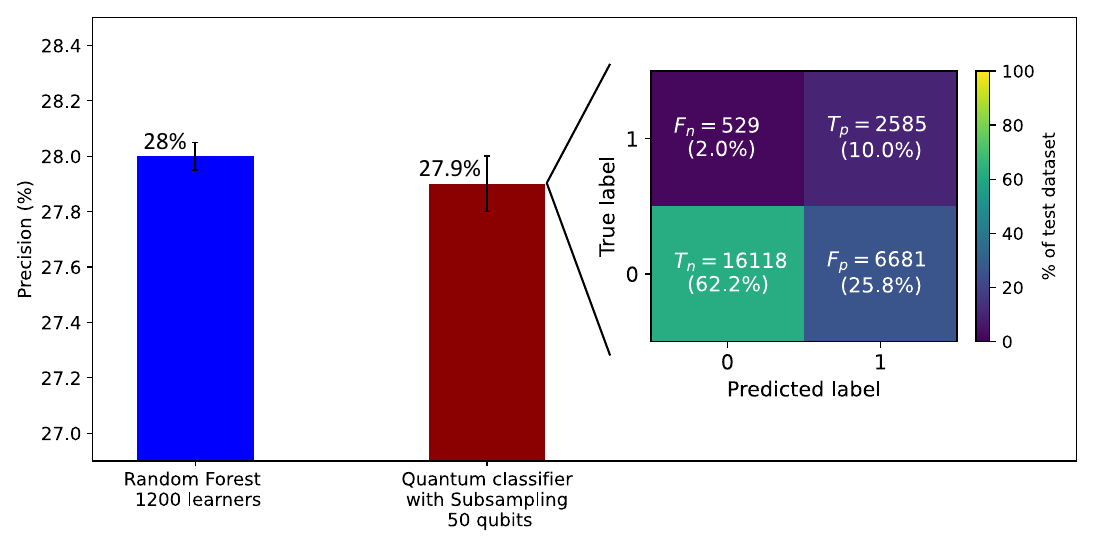}
    \caption{QPU classification results obtained using the proposed quantum classifier based on subsampling (dark red) when optimizing $50$-sized QUBOs. Its precision is compared to the one obtained with the Random Forest approach (blue) using $1200$ learners. On the right, the corresponding confusion matrix of implemented model is displayed with proportion of $T_{p/n}$ and $F_{p/n}$ as $\%$ of the test dataset.}
    \label{fig:qpu_res_50}
\end{figure*}
In this section, we present the classification results obtained using the quantum classifier based on subsampling (see Section \ref{ssec:proposed_classifier}), trained using the quantum optimizer implemented on the QPU up to $60$ qubits. This quantum classifier, leveraging the subsampling approach without boosting, is based on the optimization of QUBOs with positive off-diagonal values, amenable to efficient optimization with the current quantum hardware (see Section \ref{ssec:qubo_qpu}). We find the best results for $50$ qubits (recalling that only few statistics were available at $60$ qubits), corresponding to an initial weak ensemble of $50$ learners, whose percentages of kNNs and DTs have been optimally chosen through a hyperparameter optimization procedure. For this hyperparameter optimization (see Appendix \ref{app_sec:hyper_parameter_tuning}), the training set was split into $80\%$ training and $20\%$ cross-validation sets using stratified-shuffled splitting. \\

As depicted in Fig.\,\ref{fig:qpu_res_50}, the proposed quantum classifier is able to achieve very similar performances to the classical Random Forest algorithm. Using bitstrings with gap below $1\%$, our model reaches $P=27.9\pm0.09\%$, closely approaching the benchmark threshold $P=28.0\pm0.07\%$ for the same recall value of $R=83\%$. Very interestingly, this result is obtained with only $50$ initial learners compared to the Random Forest's ensemble of $1200$ learners. The difference in the number of learners employed is of great relevance for the interpretability of the model. Indeed, the model outputted decision for a new unseen point can be traced back more easily and better understood by the user. Further, we report the total runtimes for this model up to 50 qubits in Table \ref{tab:qpu_runtimes} of Appendix \ref{app_sec:run_times}. As seen, the best results for $50$ qubits were obtained with a total runtime of around $50$ minutes, against a total runtime of more than $3$ hours for the classical benchmark, representing a relevant practical speed-up. \\

\begin{figure*}
    \centering
    \includegraphics[width=0.9\textwidth]{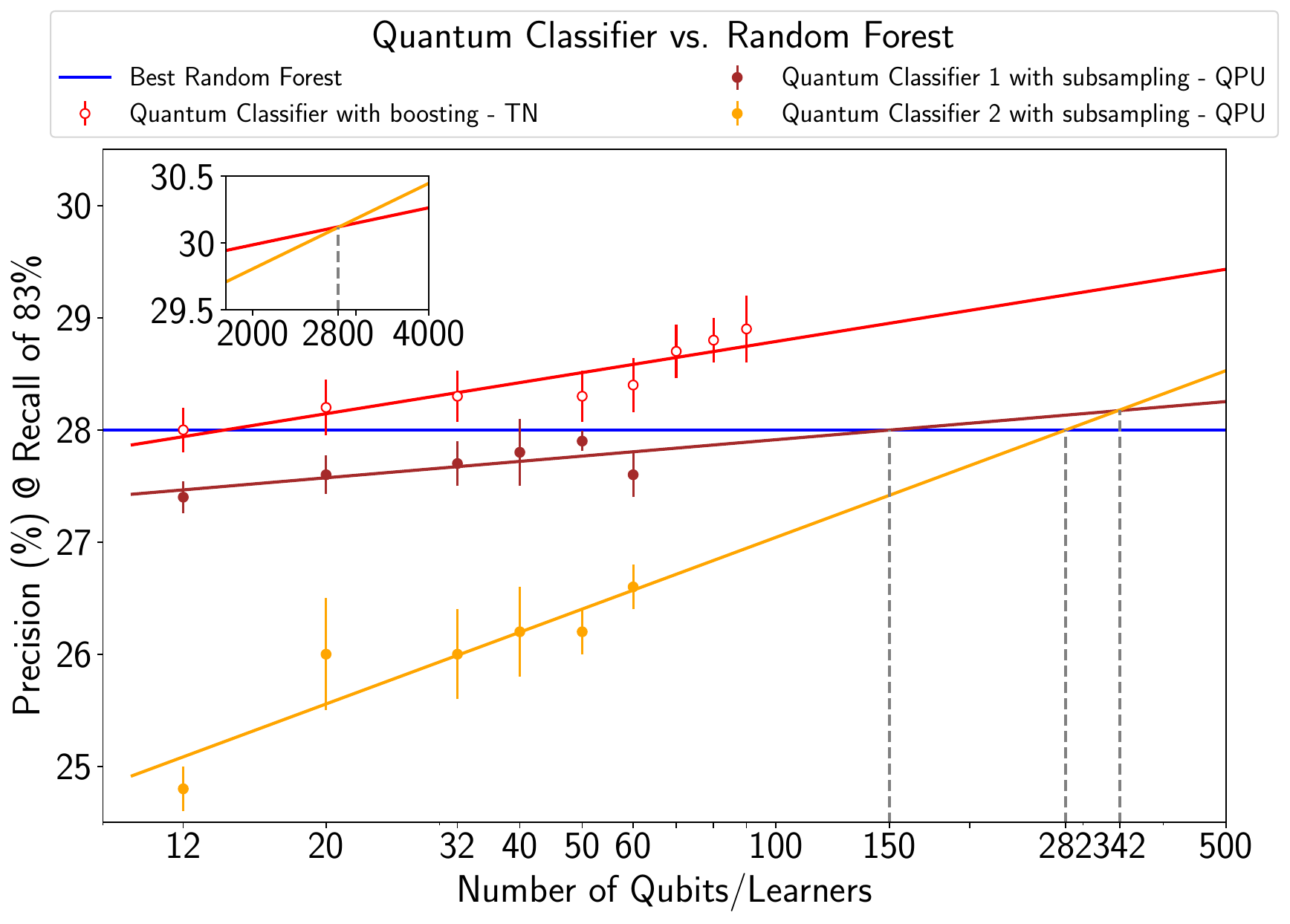}
    \caption{Scaling of precision $P$ of various proposed quantum classifiers with respect to the number of qubits, keeping $R=83\%$. The two variations of the subsampling approach (yellow, brown) are implemented on QPU (filled dot) between $12$ and $60$ qubits. The boosting approach (red) is implemented using the TN optimizer (empty dot) between $12$ and $90$ qubits. The best performance of the Random Forest classifier acts as threshold (blue). The error bars represent the variability in corresponding performance across 5 iterations/QUBOs. Scaling projections are obtained by linear extrapolation (plain line). }
    \label{fig:qc_tn_scaling}
\end{figure*}

Next, we show in Fig.\,\ref{fig:qc_tn_scaling} precision values for $R=83\%$ for two quantum classifiers (based on subsampling) with different compositions of kNNs and DTs respectively optimized to get the best possible performance up to hardware capabilities of $60$ qubits (brown curve; best results for 50 qubits are taken from here) and to get an optimal performance and a more favorable scaling trend at the same time (yellow curve). For the sake of comparing scaling trends, linear extrapolations are applied and the corresponding intersections between the interpolation lines and the benchmark threshold are marked. As seen, on the one hand, quantum classifier 1 (brown curve) presents the best performance with a slowly but increasing trend, and is expected to beat the benchmark performance at around 150 qubits. On the other hand, quantum classifier 2 (yellow line) presents lower performance in terms of available data points but an expected steeper increase, showing a predicted surpass of the benchmark (blue line) and of the quantum classifier 1 (brown line) at around $282$ and $342$ qubits, respectively.

\subsection{TN Classification Results}
In Fig.\,\ref{fig:qc_tn_scaling} we also show the mean classification performance of the quantum classifier optimized with the TN and based on boosting (red line). This model being based on the boosting procedure leverages the optimization of QUBOs with negative off-diagonal values which cannot be currently directly optimized on neutral atom QPU. It can be seen that even at low values of qubits/learners, the proposed model based on boosting, already showed the same level of performance as the Random Forest with 1200 trees. With 90 learners, it shows a mean precision score of about $P=29\%$ (reducing the false positives by $1\%$) corresponding to the recall of $83\%$. An outlook of the training and total runtimes for this heterogeneous model and the homogeneous variation with just DTs can be found in Table \ref{tab:tn_runtimes}. The best results for $90$ qubits/learners present a total runtime of the order of 20 minutes against the total runtime of more than 3 hours for the classical benchmark, attaining also in this case a relevant practical speed-up. \\

Based on the scaling projections, it can be argued that this type of model is expected to remain the best performing one. It can be seen in the inset of Fig.\,\ref{fig:qc_tn_scaling} that a crossing with the quantum classifier 2 (yellow curve) based on subsampling could occur for a large number of qubits, around 2800, although it is difficult to assess the reliability of the extrapolation for such high numbers.  

\subsection{Current Limitations and Future Upgrades}
The training of the proposed model involves optimization of a heterogeneous ensemble comprising DTs and kNNs. Due to slow execution speed and large memory requirements of kNNs, the training time of this model was found to be relatively higher than the homogeneous models comprising just DTs. In the near future, this could be overcome by using a faster implementation of kNN \cite{song2017efficient}, leveraging GPU architectures for instance.\\

Implementing the boosting variation of the classifier directly on a neutral atom hardware is at the moment hindered by several limitations. 
As presented in section \ref{ssec:qubo_qpu}, perfectly embedding a QUBO into atomic positions requires to satisfy some constraints of the form $\{Q_{ij}=U_{ij}\}_{i\neq j}$. Choosing a specific Rydberg state to implement the Ising model implies that the interaction $U$ will be positive between all atomic pairs. Choosing another Rydberg state as $\ket{1}$ can enable to have negative interactions, but only globally. Thus, QUBOs with both positive and negative off-diagonal values, such as those produced in the boosting variation, can not be natively implemented, restricting the type of classification models accessible on current neutral atom QPU. \\ 
%While it remains possible to alter a QUBO $Q\rightarrow\tilde Q$ such that $\tilde Q_{i\neqj}>0$, the resulting diagonal $Q_{ii}$ will exhibit terms with various orders of magnitude, making it difficult to resolve $\tilde Q$

However, as shown on Fig.\,\ref{fig:qc_tn_scaling}, the subsampling variation, which produces QUBOs with all positive off-diagonal values, could be able to beat the threshold set by the benchmark at around $150$ qubits. In order to optimize a heterogeneous ensemble of this size with RGS, a pattern with around $290$ traps is needed. 
In a recent work \cite{324}, a new prototype successfully produced atomic arrays of size $N=324$ with patterns of $625$ traps. As more capabilities become available for this hardware technology, future implementations may offer an opportunity to achieve an industry-relevant quantum value by beating state-of-the-art methods at larger number of qubits/learners. \\

\section*{Conclusions and Outlook}
\label{sec:conclusion}
In this paper, we propose to the best of our knowledge the first quantum-enhanced machine learning solution for the prediction of credit rating downgrades, also known as fallen-angels forecasting. Our  algorithm comprises a hybrid classical-quantum classification model based on QBoost, tested on a neutral atom quantum platform and benchmarked against Random Forest, one of the state-of-the-art classical machine learning techniques used in the Finance industry. We report that the proposed classifier trained on QPU achieved competitive performance with $27.9\%$ precision against the benchmarked $28\%$ precision for the same recall of approximately $83\%$. However, the proposed approach outperformed its classical counterpart with respect to interpretability with only $50$ learners employed versus $1200$ for the Random Forest and comparable runtimes. These results were obtained leveraging the hardware-tailored Random Graph Sampling method to optimize QUBOs up to size $60$. The RGS method showed similar performances with simulated annealing approach and was able to provide solutions to QUBO within acceptable repetitions budget.\\

We also report a classification model based on the proposed heterogeneous structure and leveraging the boosting procedure that, although is not amenable to be trained on current hardware, was trained using a quantum-inspired optimizer based on Tensor Networks. This model showed the capability to already perform better across all the relevant metrics, achieving a precision of $29\%$, $1\%$ above the benchmark, with just $90$ learners (against $1200$) and runtimes of around $20$ minutes compared to more than $3$ hours for the benchmarked Random Forest. \\

Going forward, hardware upgrades in terms of qubit numbers will lead to performance improvements. This behavior is illustrated in Fig.\,\ref{fig:qc_tn_scaling}, where we show how the precision of the quantum classifiers evolves with system size. Extrapolating from these results and keeping other factors constant, the proposed quantum classifiers could outperform the benchmarked model within a few hundred addressable qubits. In addition, hardware improvements enabling the resolution of QUBOs with negative off-diagonal values could offer additional advantages to the quantum solution and improve performance over the classical benchmark. \\

These results open up the way for quantum-enhanced machine learning solutions to a variety of similar problems that can be found in the Finance industry. Interpretability and performance improvements for real cases with complex and highly imbalanced datasets are pressing issues. As such, we foresee a large number of applications for quantum-enhanced machine learning, especially implemented on neutral atom platforms, in solving computationally challenging problems of the financial sector. 

\section*{Acknowledgments}
PASQAL thanks Mathieu Moreau, Anne-Claire Le Hénaff, Mourad Beji, Henrique Silvério Louis-Paul Henry, Thierry Lahaye, Antoine Browaeys, Christophe Jurczak, and Georges-Olivier Reymond for fruitful discussions.\\

Multiverse Computing acknowledges fruitful discussions with the rest of the technical team as well as Creative Destruction Lab, BIC-Gipuzkoa, DIPC, Ikerbasque, Diputacion de Gipuzkoa and Basque Government for constant support.\\

All the authors thank Loïc Chauvet and Ali El Hamidi.

\appendix

\section{RGS Details}
\label{app:RGS_details}
\subsection{Optimized Relabeling}
\label{app:RGS_details_opti}
\begin{figure*}
    \centering
    \includegraphics[width=0.9\textwidth]{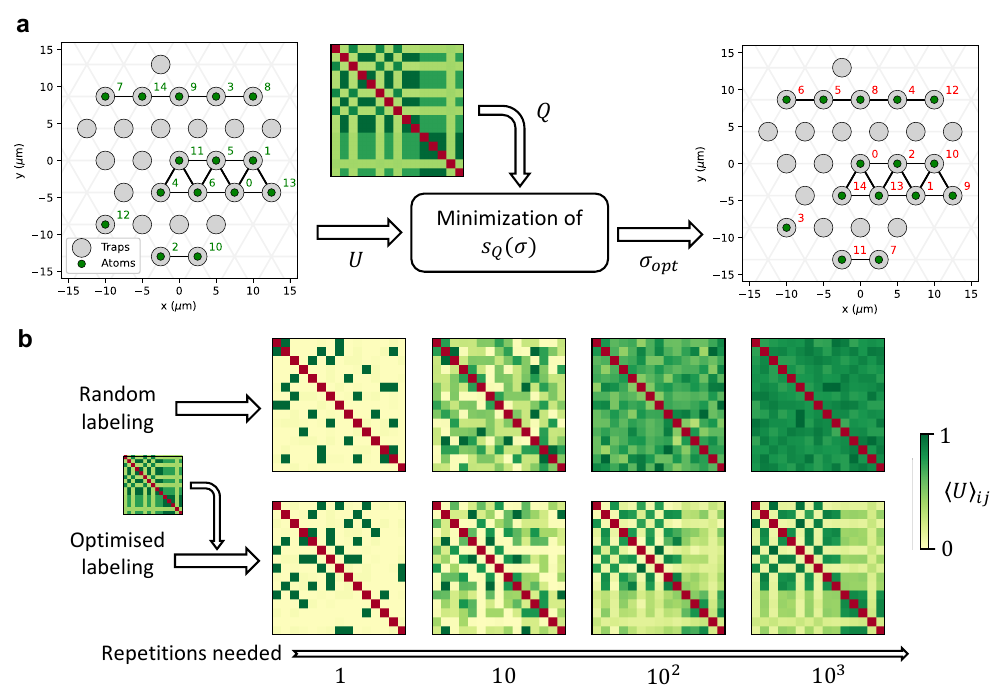}
    \caption{\textbf{a.} $15$ atoms (green dots) are filling a fraction of a triangular trap layout (gray circles). Each atom is randomly labeled (green) and from their positions an interaction matrix $U$ is derived. Using the QUBO to solve, $Q$, and Eq.\,\ref{eq:separation}, a permutation of the labels $\sigma_{opt}$ is found. The atoms are relabeled (red) such as the resulting interaction matrix replicates better $Q$. \textbf{b.} Normalized interaction matrices obtained when averaging over many repetitions of traps loading. While the random labeling (top line) produces a uniform matrix, the optimized labeling (bottom line) enables to access some features of the QUBO at each cycle, producing an average matrix resembling $Q$.}
    \label{fig:rgs_opti}
\end{figure*}
We describe here the relabeling process used in Random Graph Sampling (see Section \ref{ssec:RGS}). For a given cycle where $N$ traps out of $N_t$ are filled with atoms, a first measure of the system before the quantum processing part enables to locate the atoms, as shown in Fig.\,\ref{fig:rgs_opti}\textbf{a}. The latter are randomly labeled and this memorized labeling usually orders the bitstring measured after the quantum processing. However, we can choose another labeling more specific to the QUBO we want to solve. This post-processing step determines a labeling $\sigma_{opt}$ of the atoms such that the resulting interaction matrix better reproduces the QUBO matrix than the one obtained from the randomly generated graph. For each way of labeling the atoms from $1$ to $N$, {\it i.e.} each permutation of length $N$, we compute the separation
\begin{equation}
\label{eq:separation}
   s_Q(\sigma) = \sum_{i<j} \lVert U_{\sigma(i)\sigma(j)}-Q_{ij} \rVert,
\end{equation}
where $Q$ is the QUBO matrix, $\sigma$, a permutation of length $N$ and $U_{\sigma(i)\sigma(j)}$ the interaction term between atoms originally named $i$ and $j$. The two matrices are normalized to allow a proper comparison. We perform a random search with fixed budget $n_{iter}$ over the $N!$ possible labeling permutations. The permutation minimizing $s_Q$ is then used to read out the measured bitstring. Searching for such a permutation is reasonably fast for the sizes that can be loaded in the QPU. We have checked that for $N\leq100$, this takes less than $n_{iter}\times 2$ms. In the following, we set $n_{iter}=10N$ so as to scale only linearly with the number of qubits and not as $N!$. This may not be sufficient to identify the best permutation, but it remains enough to reproduce some of the features of the QUBO at each cycle as shown in Fig.\,\ref{fig:rgs_opti}\textbf{b}. Furthermore, on average, the whole QUBO is much better represented with this optimized relabeling step than simply using a random permutation. It is worth pointing out that this optimization step can be done retrospectively, after the quantum data has been acquired, as long as we have access to the initial traps filling. Thus, its execution time does not limit the duration of a cycle, and this can become effectively a post-processing step done on a classical computer.\\

\begin{figure*}
    \centering
    \includegraphics[width=\textwidth]{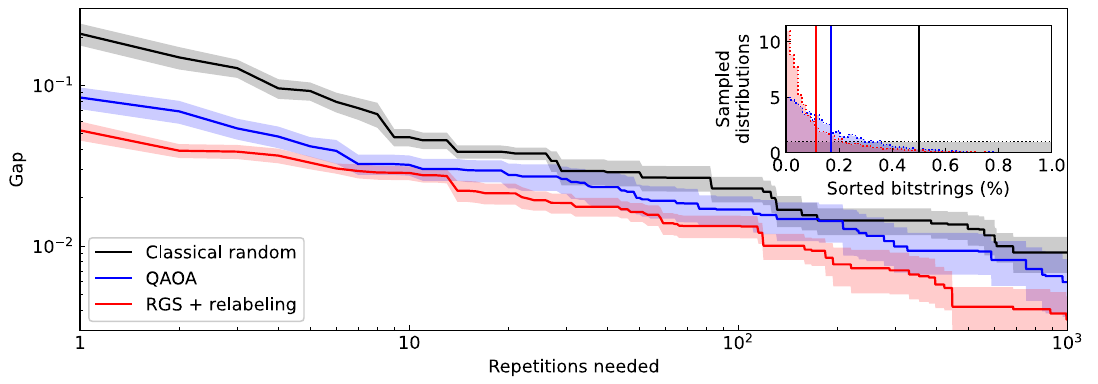}
    \caption{Results obtained from a classical random search, a numerically simulated QAOA and numerically simulated RGS with QUBO dependent labeling (averaged over $10$ random QUBO instances of size $N=15$). The inset shows the frequency and the mean of the proposed solutions with each method, ranked by their cost function value.}
    \label{fig:rgs_simu_res}
\end{figure*}

We benchmark this approach on a set of randomly generated QUBOs of size $15$ and compare it to a classical uniform sampling of $\mathbb{B}^N$ and to a numerical simulation of QAOA \cite{silverio2022pulser}, all using a similar budget of $1000$ cycles, or measurements. Getting into the detail, the QAOA algorithm is allowed $10$ iterations with $100$ cycles each in order to optimize the duration of $3$ pulses. The cost function evaluated at each iteration is $\langle \mathcal{H}_Q\rangle$ averaged over the $100$ measurements. The atoms are located at the same positions for each iteration, meaning that an experimental implementation would use the rearrangement algorithm, lengthening the duration of each cycle. In contrast, for RGS, the positions are random at each cycle while the pulse sequence remains the same, $3$ pulses with non optimized durations. We show the results of these three methods in Fig.\,\ref{fig:rgs_simu_res} with both the convergence of each one with respect to the number of cycles performed and the aggregated bitstring distributions sorted by increasing values of $\mathcal{H}_Q$. Not only does RGS converge faster, achieving a gap of less than $1\%$ with three times fewer cycles than QAOA, it also produces, on average, sampled distributions with greater concentration on bitstrings with low value. For this set, a bitstring sampled using RGS$+$relabeling is on average around the $11$ best $\%$ of $\mathbb{B}^N$ while one sampled with QAOA is around the $17$ best $\%$. 

\subsection{Configuration Clustering}
\label{app:RGS_details_clustering}
In this section, we elaborate on how to extract usable bitstrings of size $n$ from ones of size $N>n$ measured on the quantum set up. Those smaller bitstrings can in specific cases be used to solve QUBOs of size $n$.\\

\begin{figure}[!b]
    \centering
    \includegraphics[width=0.48\textwidth]{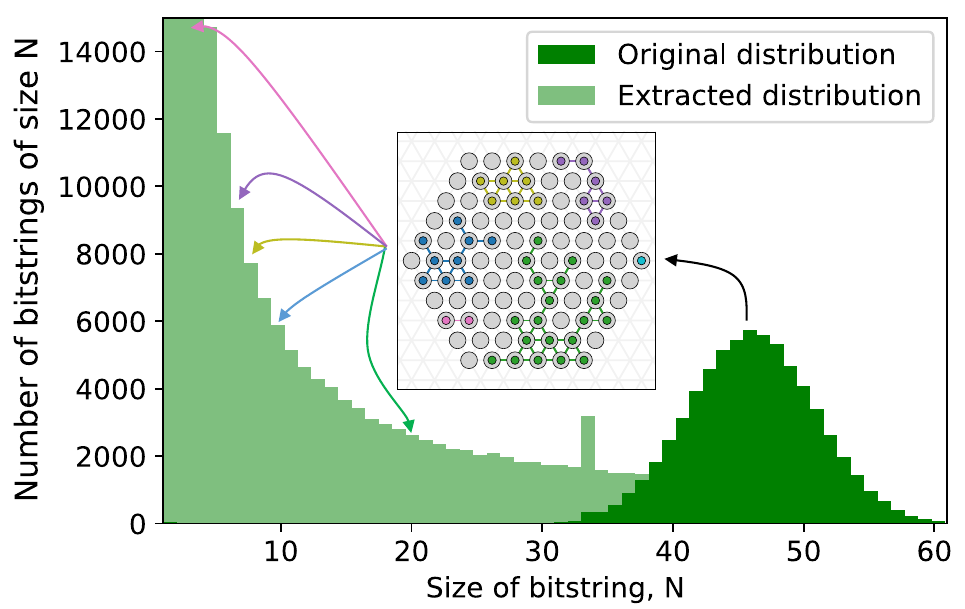}
    \caption{Clustering of atomic configurations to extract $n$-sized bitstrings from $N$-sized ones with $N>n$. From an original distribution of $65\,000$ bitstrings (dark green), we construct a larger distribution of $334\,000$ bitstrings (light green). A bitstring of size $45$ has been measured with the atomic configuration displayed in the inset. Atoms are sorted between clusters (various colors) of sizes $2,\,6,\,7,\,9,\,20$. and the initial bitstring is cut into $5$ smaller bitstrings, usable to solve QUBOs of corresponding sizes.   }
    \label{fig:subgraph}
\end{figure}

At each computation cycle, a pattern of $N_t$ traps is filled by $N\sim\mathcal{B}(N_t,p)$ atoms. Many cycles would then produce bitstrings whose sizes follow a Gaussian distribution centered around $N_t\,p$ as shown in Fig.\,\ref{fig:subgraph}. For each cycle, knowing which traps were filled (as shown in inset), we can isolate cluster of atoms with the following rule: two atoms belong to different clusters if their pair distance exceeds the pattern spacing. Therefore, due to the rapid decay of the interaction with the distance, {\it i.e.} $U(r_{ij})\propto r_{ij}^{-6}$, we can consider that clusters do not interact between them. Indeed, here, two non neighboring atoms are interacting at least $27$ times weaker than a pair of neighboring ones. Segmenting a $N$-sized bitstring leads to the extraction of $s$ smaller bitstrings with sizes $n_i$ such that $\sum_i^s n_i=N$. Applying this method to the original distribution of $\sim65\,000$ measurements, ranging in size from $34$ to $66$ atoms, outputs a wider distribution of $\sim334\,000$ bitstrings ranging in size from $1$ to $66$. This method produces bitstrings from fully interacting systems, as no atom remains isolated. However, it can reduces the number of measurements made at a large size $N$. \\

The resulting bitstrings can only be used to solve QUBOs of corresponding sizes and which would have produced the same trap pattern as the one used to acquire the original distribution. In this implementation, since we only consider QUBOs output by the subsampling approach detailed in section \ref{ssec:proposed_classifier}, all of them are similar in structure, being produced by the same dataset and with the same hyperparameters for weak leaners ensemble generation. We apply the relabeling step to the extracted distribution in order to solve the considered sets of QUBOs (see Section \ref{ssec:results_qubo_solving}) . \\ 

\section{Hyperparameter Tuning}
\label{app_sec:hyper_parameter_tuning}

For hyperparameter tuning of the proposed quantum classifier, grid-search cross-validation based optimization over a list of possible values was used. An important hyperparameter of QBoost is regularization (or $\lambda$) which serves to penalize complex models with more learners in order to achieve a better generalization on unseen data (see Eq.\,\ref{eq:cost_function}). \\  

As the number of base learners $N$ is increased, the time required to train the classifier increased. Since tuning of the regularization parameter involved re-training the classifier with many different values of $\lambda$, it was a computationally expensive procedure that needed to be sped up. Consequently, taking advantage of an insight into the cost function, it was proposed to run hyperparameter tuning procedure using a smaller and simpler variation of the classifier, comprising $N=10$ learners and use the corresponding optimal $\lambda_{10}$ for any $N>10$ by multiplying it with the factor of $10/N$, inspired by the scaling of the $\lambda$'s boundary discussed in \cite{neven1}. In other words, for all the variations of the proposed quantum classifiers, the hyperparameter tuning procedure took a fixed time of about 10 minutes (see Tables \ref{tab:qpu_runtimes} and \ref{tab:tn_runtimes}).

\section{Runtimes}
\label{app_sec:run_times}
\begin{table}[ht]
\centering
    \begin{tabular}{c|c}
    \hline
    \multicolumn{2}{c}{TOTAL TRAINING TIME (min)} \\ \hline
    \begin{tabular}[c]{c@{}}Number \\ of Qubits\end{tabular} &
    \begin{tabular}[c]{@{}c}Subsampling - QPU\\ Heterogeneous\end{tabular} \\ \hline
    12  & 31.8  \\
    20  & 35.5 \\
    32  & 37.2 \\
    40  & 43.2 \\ 
    50  & 46.5 \\
    \hline
    \end{tabular}
    \caption{Total training time including the time taken by training (ensemble training and optimization using QPU) and hyperparameter tuning (10 min) of the proposed quantum-enhanced classifier based on subsampling.}
    \label{tab:qpu_runtimes}
\end{table}
Table \ref{tab:qpu_runtimes} presents QPU runtimes of the proposed quantum classifier based on subsampling, {\it i.e.} the time taken by training which includes ensemble training on the CPU, optimization on the QPU and hyperparameter tuning (see Appendix \ref{app_sec:hyper_parameter_tuning}). In this case, the training set was over-sampled. The QPU runtimes are obtained by multiplying the number of cycles needed to output a bitstring with gap (see Eq.(\ref{eq:gap})) smaller than $1\%$ by the current repetition rate of the device. Table \ref{tab:tn_runtimes}, on the other hand, presents runtimes of the two different variations of the proposed quantum classifier based on boosting, with under-sampled training set. While the homogeneous variation is based on an ensemble of only decision trees, the heterogeneous variation comprises a mix of decision trees (DTs) and k-nearest neighbors (kNNs). It can be seen that the training of a heterogeneous classifier generally takes longer than the training of a homogeneous classifier. \\

\begin{table}[h]
\centering
\begin{tabular}{@{}c|c|c@{}}
\hline
   \multicolumn{3}{c}{TOTAL TRAINING TIME (min)} \\ \hline
  \begin{tabular}[c]{@{}c@{}}Number \\ of Qubits\end{tabular} &
  \begin{tabular}[c]{@{}c@{}}Boosting - TN\\ Homogeneous\end{tabular} &
  \begin{tabular}[c]{@{}c@{}}Boosting - TN\\ Heterogeneous\end{tabular} \\ \hline
12 & 10.1          & 11.3          \\
20 & 10.3          & 12.1          \\
32 & 10.6          & 13.6          \\
50 & 11.4          & 15.5          \\
60 & 12.1          & 16.7          \\
70 & 13.9          & 17.8          \\
80 & 14.8          & 19.4          \\
90 & 15.2          & 20            \\ \hline
\end{tabular}
\caption{Total training time including the time taken by training (ensemble training and optimization using tensor network optimization) and hyperparameter tuning (10 min) of the proposed homogeneous and heterogeneous quantum-enhanced classifiers based on boosting.}
\label{tab:tn_runtimes}
\end{table}
\bibliographystyle{unsrt}
\bibliography{references}

\end{document}